\documentclass[reprint,amsmath,amssymb,aps,superscriptaddress,twocolumn,10pt]{revtex4-1}
\usepackage[colorlinks=true,allcolors=blue]{hyperref}
\usepackage{graphicx}
\usepackage{dcolumn}
\usepackage{subfigure}
\usepackage{bm}
\usepackage{amsmath,amsfonts,amssymb}
\usepackage{acronym}
\usepackage{enumitem}
\usepackage{array}
\usepackage{multirow}
\allowdisplaybreaks
\newcommand{\TRC}{TianQin Research Center for Gravitational Physics, Sun Yat-sen University (Zhuhai Campus), 2 Daxue Rd., Zhuhai 519082, P. R. China.}
\newcommand{\SPA}{School of Physics and Astronomy, Sun Yat-sen University (Zhuhai Campus), 2 Daxue Rd., Zhuhai 519082, P. R. China.}
\begin{document}

\newcommand{\jd}[1]{\textcolor{red}{JD: #1}}
\newcommand{\pder}[2]{\frac{\partial #1}{\partial #2}}

\newacro{EMRI}{extreme mass ratio inspiral}
\newacro{MBH}{massive black hole}
\newacro{BH}{black hole}
\newacro{GW}{gravitational wave}
\newacro{AK}{analytic kludge}
\newacro{AAK}{augmented analytic kludge}
\newacro{NK}{numerical kludge}
\newacro{QAK}{quadrapole included analytic kludge}
\newacro{QNK}{quadrapole included numerical kludge}
\newacro{QAAK}{quadrapole included augmented analytic kludge}
\newacro{CO}{compact object}
\newacro{PE}{parameter estimation}
\newacro{SNR}{signal to noise ratio}
\newacro{PN}{post newtonion}
\newacro{FIM}{fisher information matrix}
\newacro{LSO}{last stable orbit}

\title{Augmented analytic kludge waveform with quadrupole moment correction}

\author{Miaoxin Liu}
\email{liumx37@mail2.sysu.edu.cn}
\affiliation{\SPA}

\author{Jian-dong Zhang}
\email{zhangjd9@mail.sysu.edu.cn}
\affiliation{\TRC}
\affiliation{\SPA}

\date{\today}
\begin{abstract}
One of the most important sources for future space-borne \ac{GW} detectors such as TianQin and LISA is \ac{EMRI}.
It happens when a stellar orgin \ac{CO} orbiting around a \ac{MBH} in the center of galaxies and has many benefits in the study of astrophysics and fundamental theories.
One of the most important objectives is to test the no-hair theorem by measuring the quadrupole moment of the \ac{MBH}.
This requires us to estimate the parameters of an \ac{EMRI} system accurately enough,
which means we also need an accurate waveform templet for this process.
Based on the fast and fiducial \ac{AAK} waveform for the standard Kerr \ac{BH},
we develop a waveform model for a metric with non-Kerr quadrupole moment.
We also analyze the accuracy of parameter estimation for different sources and detectors.
\end{abstract}

\maketitle
\section{Introduction}\label{sec:intro}

The observation of gravitational wave has provided a new approach to probe the universe.
By analyzing the signal of GW150914 \cite{Abbott:2016blz} and all the \ac{GW} events observed by ground-based observatory,
the constraint for the theory of gravity has been enhanced to a higher  \cite{TheLIGOScientific:2016src,LIGOScientific:2019fpa}.
For example, the mass of graviton should be less than $\sim10^{-23}eV$.
However, several space-borne \ac{GW} detectors is proposed to be launched in the 2030s,
such as the heliocentric detector LISA \cite{Audley:2017drz} and the geocentric detector TianQin \cite{Luo:2015ght}.

Different from the ground-based \ac{GW} detectors which are sensitive to the hundred Hz \ac{GW} signals,
space-borne detectors are sensitive to the micro Hz band.
Among all the sources in this band, \ac{EMRI} is one of the most important targets \cite{Babak:2017tow,Fan:2020zhy}.
It happens when a stellar mass compact object,
which could be a neutron star or a black hole,
captured by the \ac{MBH} in the center of a galaxy,
and orbiting around the \ac{MBH} in the near horizon region for more than thousands cycles before plunges into the \ac{MBH}.
Since the \ac{CO} will stay in the strong gravity region for a long time,
so the \ac{GW} radiated by \ac{EMRI} will carry a wealth of information about the geometry and environment around the \ac{MBH}.
Then by analyzing the \ac{GW} signal emitted by an \ac{EMRI} system,
one can verify the existence of different kinds of dark matters surrounding the \ac{MBH}
\cite{Hannuksela:2019vip,Hannuksela:2018izj} or test the ``no-hair'' nature \cite{Ryan:1997hg,Cardoso:2016ryw} of the \ac{MBH}.

In this work, we will focus on the issue of no-hair theorem.
More explicitly, one of the performance of no-hair theorem is that the multipole moments of a Kerr \ac{BH} in GR
is completely determined by its mass and spin as $M_l+iS_l=M(ia)^l$ \cite{Hansen:1974zz,Thorne:1980ru,Fodor1989Multipole},
where the mass multipole moments $M_l$ and mass-current multipole moments $S_l$ are real numbers.
But in other theories of gravity or other \ac{BH} solutions,
it will also be influenced by the additional parameters.
So by measuring the mass, spin, and quadrupole moment of a \ac{BH},
and check whether they satisfy this relation within the range of error,
we can judge whether it's a Kerr \ac{BH} or not.
This is actually determined by the precision of \ac{PE} for those parameters.
Then an accurate waveform including this effect is needed to enhance the ability of testing no-hair theorem.

The waveform for an \ac{EMRI} system is very complex,
since many higher order perturbation effects will play important role in orbit evolution and \ac{GW} generation.
Due to the extreme mass ratio, inspiral will take a very long time, up to several years ($\sim10^8$ seconds) or equivalently $\sim 10^5$ cycles.
Then the longer the dephasing time for the waveform, the less the segments needed in the semi-coherent detection.
And the critical \ac{SNR} will be lower.
On the other hand, in the semi-coherent search, we need to generate a large amount of waveform templates,
so the time of generation is also very important.

In spite of the Teukolsky-based waveform based on the black hole perturbation theory which is computationally expensive,
the kludge family is a class of very important and widely used methods
which can be generated quickly and capture the main features of the true signals.
The basic idea of kludge is to combine different features of orbital evolution and \ac{GW} emission directly
without the consideration of their coupling.
Roughly speaking, there are three kludge models.
The \ac{AK} model \cite{BC2004} is constructed by calculating the orbit evolution with \ac{PN} expansion
under the consideration of Lense-Thirring precession and pericenter precession.
Then the waveform is generated with Peter-Mathews formula \cite{Peters:1963ux,Peters:1964gr} in the quadrupole approximation.
\ac{AK} generates the waveform very fast, but the accuracy is limited by the kludge method.
However, we can improve its precision by simply adding higher order terms.
So it's still widely used in a lot of order of magnitude analyses.
The \ac{NK} model \cite{GG2006} provides a more accurate waveform with a slightly expensive computational cost.
It calculates the trajectory evolution first in the phase space defined by the constants,
and integrates out the more reality trajectory in the coordinate space.
Then the \ac{GW} waveform can be calculated with the leading order quadrupole approximation.
In recent years, the \ac{AAK} model \cite{Chua:2017ujo,Chua:2016jnd} is also developed to combine the advantages of the previous models.
It maps the parameters of \ac{AK} waveform to match the frequencies of \ac{NK} waveform,
and then uses the new parameters to generate the waveform with \ac{AK}.
Generally, \ac{AAK} shows an excellent overlap with \ac{NK}, retaining the speed advantage of \ac{AK}.

By adding the quadrupole moment term to the \ac{PN} orbit evolution equation of \ac{AK},
LISA's ability of measuring the quadrupole of \ac{MBH} has been studied \cite{Barack:2006pq}.
This waveform model is denoted as \ac{QAK} in this paper since it can produce the waveform for
an \ac{EMRI} system whose central massive object could possess arbitrary quadrupole moment.
However, although the \ac{FIM} method is just an order of magnitude estimation,
but with a more accurate waveform we can get a better estimation.

Beside considering to include the quadrupole moment corrections,
there also exist many other alternative methods using the \ac{GW} of EMRIs to test the nature of gravitational theory and black hole.
In \cite{Vigeland:2011ji}, by requiring the existence of a perturbative second-order Killing tensor for the bumpy black hole,
the three constants for motion are still possessed in the parametric deformed Kerr metric for non-GR deviations.
Then the leading order bump corrections to \ac{AK} waveforms are obtained by \cite{Gair:2011ym}.
This work applies the ppE framework into the EMRI waveform computations,
and push forward a first attempt toward complete and model-independent tests of General Relativity with \ac{EMRI}.
The corresponding FIM analysis is also taken in \cite{Moore:2017lxy}.
Another very important progress is the development of a framework
for testing GR with EMRI observations in \cite{Chua:2018yng}.
The Bayesian method is used in the analysis using the bumpy AK waveform in \cite{Gair:2011ym}.

In this paper, we describe a \ac{QAAK} waveform model based on the more accurate \ac{AAK} model.
(The code of QAAK is developed based on the AAK code from the EMRI Kludge Suite, which can be found from the following url: https://github.com/alvincjk/EMRI\_Kludge\_Suite.)
We first update the \ac{NK} waveform to include the quadrupole moment,
and then map the parameters of \ac{QAK} to match the frequencies.
We also calculate the \ac{PE} result based on the \ac{QAAK} waveform.

A brief overview of the kludge waveforms is given in \ref{sec:kludge}.
Then we review the quadrupole correction in the \ac{QAK} waveform in \ref{sec:quadrupole},
and present the correction we used in the \ac{QAAK} waveform model in \ref{sec:correction}.
Finally we analyze the accuracy of parameter estimation for various sources and detectors in \ref{sec:estimation}.
Then the paper ends with a conclusion in \ref{sec:conclusion}.
We use the geometric unit with $G=c=1$.

\section{a brief review of the kludge family}\label{sec:kludge}

Currently, there are three members in the kludge family which is used for the generation of \ac{EMRI}'s waveform,
they are \ac{AK} \cite{BC2004}, \ac{NK} \cite{BEA2007} and \ac{AAK} \cite{Chua:2017ujo,Chua:2016jnd}.
However, there exist many other \ac{EMRI} waveform models,
such as \cite{ST2003,FEH2016} and \cite{vandeMeent:2018rms} which include the self-force correction, and so on.
But we will not discuss these models here.

Generally, the kludge family described the inspiral waveform for a \ac{CO} which is regarded as a point particle with mass $\mu$,
in the background of a Kerr \ac{BH} with mass $M$ and spin $a$.
The metric is written in the Boyer-Lindquist coordinates:
\begin{equation}
\label{eq:Kerr}
\begin{split}
ds^2=&-\left(1-\frac{2Mr}{\Sigma}\right)dt^2
-\frac{4aMr\sin^2\theta}{\Sigma}dtd\phi\\
&+\left(\Delta+\frac{2Mr(r^2+a^2)}{\Sigma}\right)\sin^2\theta d\phi^2\\
&+\frac{\Sigma}{\Delta}dr^2+\Sigma d\theta^2,
\end{split}
\end{equation}
with
\begin{equation}
\begin{split}
\Delta=r^2-2Mr+a^2,~~~\Sigma=r^2+a^2\cos^2\theta.
\end{split}
\end{equation}
And according to the definition of \ac{EMRI}, we have $M\gg\mu$.
In fact, all the masses here are red-shifted mass $M(1+z)$,
but we will not use a subscript to distinguish these variables.
The direction of the spin for the \ac{MBH} is represented by the unit vector $\hat{S}$,
or equivalently by $\theta_K$ and $\phi_K$.

In the kludge family, orbit is considered as an eccentric and non-equatorial one.
We will use $e$ and $p$ to represent eccentricity and semi-latus rectum.
The pericenter and apocenter distance is $r_p=p/(1+e)$ and $r_a=p/(1-e)$.
The direction of the \ac{CO}'s orbital angular momentum is represented by $\hat{L}$.
Then the angle between $\hat{L}$ and $\hat{S}$ is $\iota$,
the azimuthal of $\hat{L}$ is $\alpha$,
and the angle between $\hat{L}\times\hat{S}$ and the pericenter is $\tilde{\gamma}$.
For the \ac{CO}'s motion, $\Phi$ is the mean anomaly, and $\nu$ is the radial frequency.

Apart from the geodesic parameters, the orbit can also be described by the conserved quantities:
the orbital energy $E$, the angular momentum on the direction of $\hat{S}$ which is $L_z$,
and the Carter constant $K$.
Equivalently, it can be described by the dimensionless fundamental angular frequencies corresponding to the three spacial coordinate:
$\omega_r,~\omega_\theta,~\omega_\phi$.

The location of the source is defined by the angular position $(\theta_S,\phi_S)$ and the luminosity distance $D_L$.
In fact, in this paper we consider a sky-averanged response by the \ac{GW} observatory.
So the position of the source will not appear in the following discussion,
since it will only influence the antenna patten function,
and has nothing to do with the waveform generation.

Then by assuming the initial value and equation of motion of these parameters,
the orbit evolution and then the \ac{GW} waveform can be produced by using the following three different kludge models.

\subsection{Analytic kludge}\label{subsec:AK}

In AK model, the orbital evolution is given by five first order ordinary differential equations
of $(\Phi,\nu,\tilde{\gamma},e,\alpha)$.
The equations are given by the \ac{PN} method, and is presented by Eqs. (27)--(31) in \cite{BC2004}.
In that paper, the equations of $\nu$ and $e$ are accurately through 3.5 \ac{PN} order,
and the equations of $\tilde{\gamma}$ and $\alpha$ are accurately through 2 \ac{PN} order,
while they are all accurately through order 1 for the spin $a$.
Obviously, higher order terms can be added directly into these equations if available.

Then by integrating out these geodesic parameters,
we can obtain the waveform with n-harmonics by using the Peter-Matthews method in the quadrupole approximation,
which is described by Eqs. (7)--(10) in \cite{BC2004}.

As a result of computational efficiency, \ac{AK} is used in various works of the science case study of \ac{EMRI}
\cite{Gair:2004iv,Babak:2017tow,Barack:2006pq,Fan:2020zhy}.
It has also been used in the mock LISA data challenges for the generation of
injected signals and templates for search \cite{BEA2008,BEA2008a,BEA2010,BGP2009}.
However, the insufficient accuracy will reduce the performance of detection and \ac{PE}
if it's applied to analyze the data sets containing realistic EMRI signals.
But for a \ac{PE} analyze based on \ac{FIM}, it will be accurate enough for an order of magnitude estimation
of the \ac{EMRI} signals with sufficiently high \ac{SNR}.

\subsection{Numerical kludge}\label{subsec:NK}

In the \ac{NK} model, the orbit is given by integrating the geodesic equations:
\begin{equation}\label{eq:geodesic}
\begin{split}
\Sigma\frac{dr}{d\tau}&=\pm\sqrt{V_r},\\
\Sigma\frac{d\theta}{d\tau}&=\pm\sqrt{V_\theta},\\
\Sigma\frac{d\phi}{d\tau}&=V_\phi,\\
\Sigma\frac{dt}{d\tau}&=V_t.
\end{split}
\end{equation}
where $\tau$ denotes the proper time,
and the potentials $V_{t,\theta,\phi,t}$ are functions of the constants $(E,L_z,K)$ and the coordinates $(r,\theta)$.

For a bound orbit, the trajectory is determined by $(r_a,r_p,\theta_{min})$, where $\theta_{min}$ is the minimum value of $\theta$.
In fact, $r_a$ and $r_p$ are the roots of $V_r$, while $\theta_{min}$ is the smaller root of $V_\theta$.
Equivalently, we can also describe a trajectory with $p=\frac{2r_ar_p}{r_a+r_p}$,
$e=\frac{r_a-r_p}{r_A+r_p}$, and $\iota=\frac{\pi}{2}-\theta_{min}$.
According to the well-known \ac{PN} result,
the time derivatives of the constants $(\dot E,\dot L_z,\dot K)$ are functions of $(M,a,\mu)$ and $(p,e,\iota)$ \cite{Gair:2005ih}.
Then the evolution of the constants can be integrated out.
And the trajectory of the \ac{CO} can be calculated out afterwards.
So the waveform can be obtained from the inspiral trajectory.

The accuracy of \ac{NK} waveform is well enough to agree with the Teukolsky-based waveform, which is much better than \ac{AK}.
But the computation cost is also more expensive, since it needs to integrate the trajectory both in the phase space and the coordinate space elaborately.

\subsection{Augmented analytic kludge}\label{subsec:AAK}

The \ac{AAK} model possesses both the speed of \ac{AK} and the accuracy  of \ac{NK}.
It first generates a small section of trajectory with \ac{NK},
and then maps the \ac{AK} trajectory to the \ac{NK} result and finds out the best-fit parameters.
Then the waveform will be generated by \ac{AK} with these new parameters.

Briefly speaking, given the orbit evolution in \ac{NK},
by defining a timelike parameter $\lambda=\int d\tau/\Sigma$,
we can define the dimensionless fundamental frequencies $\omega_{r,\theta,\phi}$ as
\begin{equation}\label{eq:funfre}
\begin{split}
&\omega_r=\frac{2\pi}{M\Lambda_r\Gamma},\hspace{10ex}
\omega_\theta=\frac{2\pi}{M\Lambda_\theta\Gamma},\\
&\omega_\phi=\frac{1}{M\Lambda_r\Lambda_\theta\Gamma}
\int_0^{\Lambda_r}d\lambda_r\int_0^{\Lambda_\theta}d\lambda_\theta V_\phi,
\end{split}
\end{equation}
with $\Lambda_r$, $\Lambda_\theta$, and $\Gamma$ are given by
\begin{equation}\label{eq:lambda}
\begin{split}
&\Lambda_r=2\int_{r_p}^{r_a}\frac{dr}{\sqrt{V_r}},~~~
\Lambda_\theta=2\int_{\theta_{min}}^{\frac{\pi}{2}}\frac{d\theta}{\sqrt{V_\theta}},\\
&\Gamma=\frac{1}{\Lambda_r\Lambda_\theta}
\int_0^{\Lambda_r}d\lambda_r\int_0^{\Lambda_\theta}d\lambda_\theta V_t.
\end{split}
\end{equation}

As a function of $(M,a,p)$, these fundamental frequencies can be related to the orbital frequencies as
\begin{equation}\label{eq:map}
\begin{split}
\dot{\Phi}(\tilde{M},\tilde{a},\tilde{p})&=\omega_r(M,a,p),\\
\dot{\gamma}(\tilde{M},\tilde{a},\tilde{p})&=\omega_\theta(M,a,p)-\omega_r(M,a,p),\\
\dot{\alpha}(\tilde{M},\tilde{a},\tilde{p})&=\omega_\phi(M,a,p)-\omega_\theta(M,a,p).
\end{split}
\end{equation}
The left hand side is given by the \ac{AK} orbital equations.
By solving these equations, we can get the unphysical parameters $(\tilde{M},\tilde{a},\tilde{p})$.
Next, the waveform can be generated by \ac{AK} with the new parameters.

To reduce the computational cost, the map is done on a small section.
Then, the correction along the local trajectory will be extrapolated to global inspiral as fitted polynomials.
The details can be found in \cite{Chua:2017ujo}.
The \ac{AK} part in \ac{AAK} is replaced by a higher order equation in \cite{SF2015}.

On the other hand, the \ac{LSO} cutoff is also different from the one used in \ac{AK} for Schwarzschild and Kerr.
The plunge happens when
\begin{equation}\label{eq:plunge}
\begin{split}
\frac{\partial^2V_r(r,a,E,L_z,K)}{\partial r^2}&\leq
\frac{\partial V_r(r,a,E,L_z,K)}{\partial r}\\
&=V_r(r,a,E,L_z,K)=0
\end{split}
\end{equation}
In practice, AAK uses Kepler's third law to estimate $p$ by frequency roughly, then check the stability of $(e,\iota,p)$.

\section{quadrupole moment and its influence on the EMRI waveform}\label{sec:quadrupole}

The famous ``no hair'' theorem \cite{Cardoso:2016ryw} of General Relativity has an important prediction:
A black hole will ``settles down'' to the Kerr solution almost immediately after its formation,
and all of its property can be totally expressed in terms of two physical parameters alone: its mass $M$ and spin parameter $a$.

As a consequence of these theorem, the multipole moments of a Kerr \ac{BH} is characterized by only $M$ and $a$ according to the neat relation \cite{Hansen:1974zz,Thorne:1980ru,Fodor1989Multipole}:
\begin{equation}\label{moments}
M_l+iS_l=M(ia)^l,
\end{equation}
where $M_l$ and $S_l$ are the mass and mass-current multipole moments, respectively, and $a=S/M$ is the spin parameter.
For instance, the quadrupole moment $\mathcal{Q} \equiv M_2$ of the pure Kerr geometry is given by
\begin{equation}\label{KerrQ}
\mathcal{Q} = - S^2/M.
\end{equation}
But for other \ac{BH} solutions in other theories of gravity, the relation may be modified, such as  \cite{Pappas:2014gca} in the scalar-tensor theory and \cite{Vigeland:2010xe} for bumpy black holes which we will discuss later.
For simplify, we will use the dimensionless quadrupole as $Q = \mathcal{Q}/M^3$ in our calculation.
So by measuring the value of the quadrupole moment, we can study the nature of the \ac{BH}.
Then we first need to construct a waveform model including the effect of quadrupole moment corrections.

The situation is more complicated if we consider a non-Kerr spacetime with quadrupole deviate from Kerr value.
In the \ac{QAK} waveform given by \cite{Barack:2006pq} with the lowest order corrections for $\mathcal{Q}$,
the related terms in the equations for $\tilde{\gamma}$ and $\alpha$ is taken from \cite{Lai:1995hi},
while the terms in the equation of $\nu$ is obtained by replacing the terms quadratic in the spin parameter.

To obtain a \ac{QAAK} model, we need both an enhanced \ac{QNK} model with arbitrary quadrupole moment,
and an enhanced \ac{QAK} model with higher order terms suitable with the one in \ac{AAK} model.
Then the following procedure will be done almost the same as the original \ac{AAK} model.
In the parameters' mapping of \ac{QAAK}, we didn't include $Q$ since it's a higher order correction.

For the enhanced \ac{QAK} model with higher order corrections,
we choose a rough approach by replacing all the terms quadratic in spin with $-Q$ in the higher order equations \cite{SF2015}.
The same operation is applied to the evolution of the constants in the enhanced \ac{QNK} model given by \ac{PN}.
So $(\dot E,\dot L_z,\dot K)$ are now functions of $(M,a,Q,\mu)$ and $(p,e,\iota)$.
Then the final step is to obtain the fundamental frequencies for the metric with the corresponding quadrupole corrections.
And then the unphysical parameters $(\tilde{M},\tilde{a},\tilde{p})$ can be obtained by a direct mapping.
We should notice that in the mapping of parameters for \ac{QAAK}, we will keep $Q$  fixed.
So the solutions of the frequencies equations will have a slightly different with the \ac{AAK} result, since the quadratic terms of $S$ are now fixed terms of $Q$.

Note that in \ac{QNK} model, we merely evolve the $(E,L_z,K)$ and calculate the Kerr frequencies with them.
The geodesic equation is not modified, since $K$ is not well defined in a spacetime with arbitrary quadrupole moment.
The quadrupole correction is not included in the convert $(e,\iota,p)\to(E,L_z,K)$ and the check of plunge either.

In general, the most important thing we need to do is to obtain the frequency correction corresponding to the variation of quadrupole moment.
We find that this has been obtained for the bumpy Kerr black hole with a quadrupole bump \cite{Vigeland:2009pr}.

\section{The calculation of frequency correction}\label{sec:correction}

Motivited by probing the multipole moments deviation, the bumpy black hole which can deviates in a small, controllable manner from the exact black holes of GR is introduced in \cite{Collins:2004ex} for the Schwarzschild case.
Then the bumpy Kerr is first obtained by \cite{Glampedakis:2005cf}, and then obtained by \cite{Vigeland:2009pr} using the Newman-Janis algorithm\cite{nj}.
The bumpy Kerr metric is given by $g_{\mu\nu} = {\hat g}_{\mu\nu} + b_{\mu\nu}$,
where the traditional Kerr part \eqref{eq:Kerr} is ${\hat g}_{\alpha\beta}$,
and the bumpy part $b_{\mu\nu}$ is:
\begin{equation}\label{eq:bumpy_metric}
\begin{split}
b_{tt} =&-2\left(1-\frac{2Mr}{\Sigma}\right)\psi_1,\\
b_{rr} =&2\left(\gamma_1-\psi_1\right)\frac{\Sigma}{\Delta},\\
b_{\phi\phi} =&\Delta\sin^2\theta\left[ \left(\gamma_1-\psi_1\right)
\frac{8a^2M^2r^2\sin^2\theta}{\Delta\Sigma(\Sigma-2Mr)}\right.\\
&\left.-2\psi_1\left(1-\frac{2Mr}{\Sigma}\right)^{-1} \right],\\
b_{\theta\theta} =&2\left(\gamma_1-\psi_1\right)\Sigma,\\
b_{tr} =&-\gamma_1\frac{2a^2Mr\sin^2\theta}{\Delta\Sigma},\\
b_{t\phi} =&\left(\gamma_1-2\psi_1\right)
\frac{2aMr\sin^2\theta}{\Sigma},\\
b_{r\phi} =& \gamma_1a\sin^2\theta\\
&\left[\left(1-\frac{2Mr}{\Sigma}\right)^{-1}
-\frac{4a^2M^2r^2\sin^2\theta}{\Delta\Sigma(\Sigma-2Mr)}\right].
\end{split}
\end{equation}

In the case of quadrupole bumps, the $\psi_1$ and $\gamma_1$ is given by:
\begin{equation}
\begin{split}
&\psi_1^{l=2}(r, \theta) = \frac{B_2M^3}{4}\sqrt{\frac{5}{\pi}}\frac{1}{d(r,\theta,a)^3}\\
&\left[\frac{3L(r,\theta,a)^2\cos^2\theta}{d(r,\theta,a)^2}- 1\right],\\
&\gamma_1^{l=2}(r, \theta) = B_2\sqrt{\frac{5}{\pi}} \left[\frac{L(r,\theta,a)}{2}\right.\\
&\left.\frac{c_{20}(r,a) +c_{22}(r,a)\cos^2\theta +c_{24}(r,a)\cos^4\theta}{d(r,\theta,a)^5} - 1\right],
\end{split}
\end{equation}
where
\begin{equation}
\begin{split}
d(r, \theta, a) &= \sqrt{r^2 - 2Mr + (M^2 + a^2)\cos^2\theta},\\
L(r,\theta,a) &= \sqrt{(r - M)^2 + a^2\cos^2\theta},
\end{split}
\end{equation}
and
\begin{equation}
\begin{split}
c_{20}(r,a) =& 2(r-M)^4 - 5M^2(r-M)^2 + 3M^4,\\
c_{22}(r,a) =& 5M^2(r-M)^2 - 3M^4 +\\
&a^2\left[4(r-M)^2 - 5M^2\right],\\
c_{24}(r,a) =& a^2(2a^2 + 5M^2).
\end{split}
\end{equation}
Then the quadrupole moment is given by
\begin{equation}\label{bumpyQ}
\mathcal{Q} =-Ma^2-B_2M^3\sqrt{5/4\pi}=\mathcal{Q}_K+\Delta\mathcal{Q}.
\end{equation}
or equivalently, $\Delta Q=-B_2\sqrt{5/4\pi}$.

For a point particle moving on the bumpy Kerr with mass $\mu$ and momentum $p^\mu$,
the Hamiltonian ${\cal H}$ is given by:
\begin{equation}
\mathcal{H} =\frac{1}{2}g^{\alpha\beta} p_\alpha p_\beta = -\frac{\mu^2}{2}=\hat{\mathcal{H}} + \mathcal{H}_1\;,
\label{eq:bumpy_hamiltonian}
\end{equation}
where $\hat{\mathcal{H}}$ is the Hamiltonian corresponding to the Kerr background,
and $\mathcal{H}_1$ represents the influence of the spacetime's bumpiness.

Then the orbital frequencies of Kerr is given by
\begin{equation}
\mu\hat \Omega^\mu = \frac{\partial\hat{\mathcal{H}}}{\partial{\hat J}_\mu}
\label{eq:gen_kerr}
\end{equation}
where the derivatives are taken with respect to the action variables defined for the background motion:
\begin{equation}
\hat J_i \equiv \frac{1}{2\pi} \oint p_idx^i,~~~\hat J_t \equiv -E.
\label{eq:J}
\end{equation}
And the frequencies' shift can be expressed by averaged $\mathcal{H}_1$:
\begin{equation}
\mu\delta\Omega^\mu = \frac{\partial\langle\mathcal{H}_1\rangle}{\partial{\hat J}_\mu},
\label{eq:gen_shifts}
\end{equation}
while the orbit averaged form of $\mathcal{H}_1$ can be defined as:
\begin{equation}
\langle\mathcal{H}_1\rangle = \frac{1}{\Gamma\Lambda_{r}\Lambda_{\theta}}\int_0^{\Lambda_r}
d\lambda_r \int_0^{\Lambda_\theta} d\lambda_{\theta}\,\mathcal{H}_1 V_t,
\label{eq:H1_averaged}
\end{equation}

Once the derivatives of $\mathcal{H}$ and background frequencies are available, it is simple to compute the changes to the observable frequencies. Expanding
\begin{equation}
\omega^i = \frac{\Omega^i}{\Omega^t} =\frac{\hat\Omega^i + \delta\Omega^i}{\hat\Omega^t+ \delta\Omega^t}\equiv\hat\omega^i + \delta\omega^i,
\end{equation}
So the deviation of frequencies can be written as
\begin{equation}
\delta\omega^i = \frac{\delta\Omega^i}{\hat\Omega^t} -
\frac{\hat\omega^i\,\delta\Omega^t}{\hat\Omega^t}\;,
\end{equation}

Then by replacing all the parameter $B_2$ with $-2\Delta Q\sqrt{\pi/5}$,
we can get the fundamental frequencies corresponding to the spacetime with quadrupole moment deviation $\Delta Q$.
However, in the Newtonian limit, the deviation is given by
\begin{equation}
\begin{split}
\delta\omega^r=&-\frac{3\Delta Q}{4M}\frac{1}{p^{7/2}}(1-e^2)^2(2\sin^2\theta_m-1),\\
\delta\omega^\theta=&-\frac{3\Delta Q}{4M}\frac{1}{p^{7/2}}(1-e^2)^{3/2}\\
&[\sin^2\theta_m(5+3\sqrt{1-e^2})-\sqrt{1-e^2}-1],\\
\delta\omega^\phi=&-\frac{3\Delta Q}{4M}\frac{1}{p^{7/2}}(1-e^2)^{3/2}[\sin^2\theta_m\\
&(5+3\sqrt{1-e^2})-2\sin\theta_m-\sqrt{1-e^2}-1].
\end{split}
\end{equation}

\section{Parameter estimation result}\label{sec:estimation}

Given the quadrupole moment included waveform and a specific detector,
we can get the expected accuracy of \ac{PE} with the \ac{FIM} method.
The inner product is defined as \cite{CF1994}:
\begin{equation}
\langle a|b\rangle=2\int_0^\infty df\,\frac{\tilde{a}^*(f)\tilde{b}(f)+\tilde{a}(f)\tilde{b}^*(f)}{S_n(f)}.
\end{equation}
The \ac{SNR} is defined as $\rho=\sqrt{\langle h|h\rangle}$,
and the \ac{FIM} is
\begin{equation}
\Gamma_{ij}=\left\langle\frac{\partial h}{\partial\lambda_i}\big|\frac{\partial h}{\partial\lambda_j}\right\rangle,
\end{equation}
where $\lambda_i$ are parameters which are used to generate the waveform.
When the \ac{SNR} of the signal is high enough, then the \ac{PE} accuracy is given by:
\begin{equation}
\delta\lambda_i\approx\sqrt{(\Gamma^{-1})_{ii}}.
\end{equation}

\begin{figure*}
\centering
\includegraphics[width=0.32\textwidth]{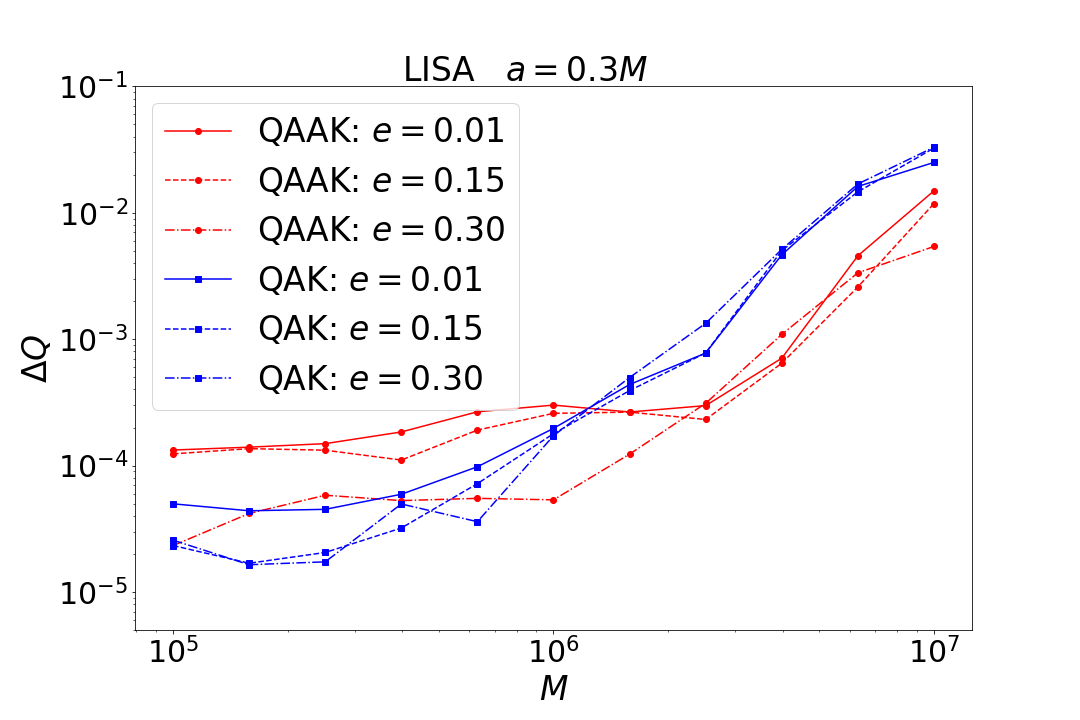}
\includegraphics[width=0.32\textwidth]{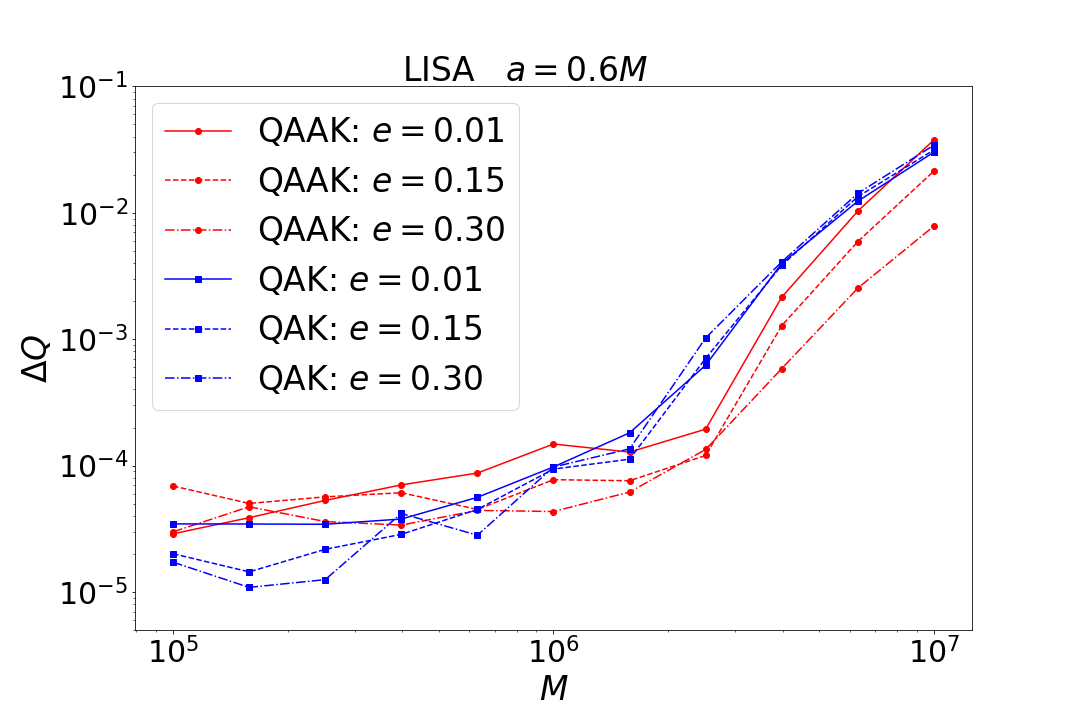}
\includegraphics[width=0.32\textwidth]{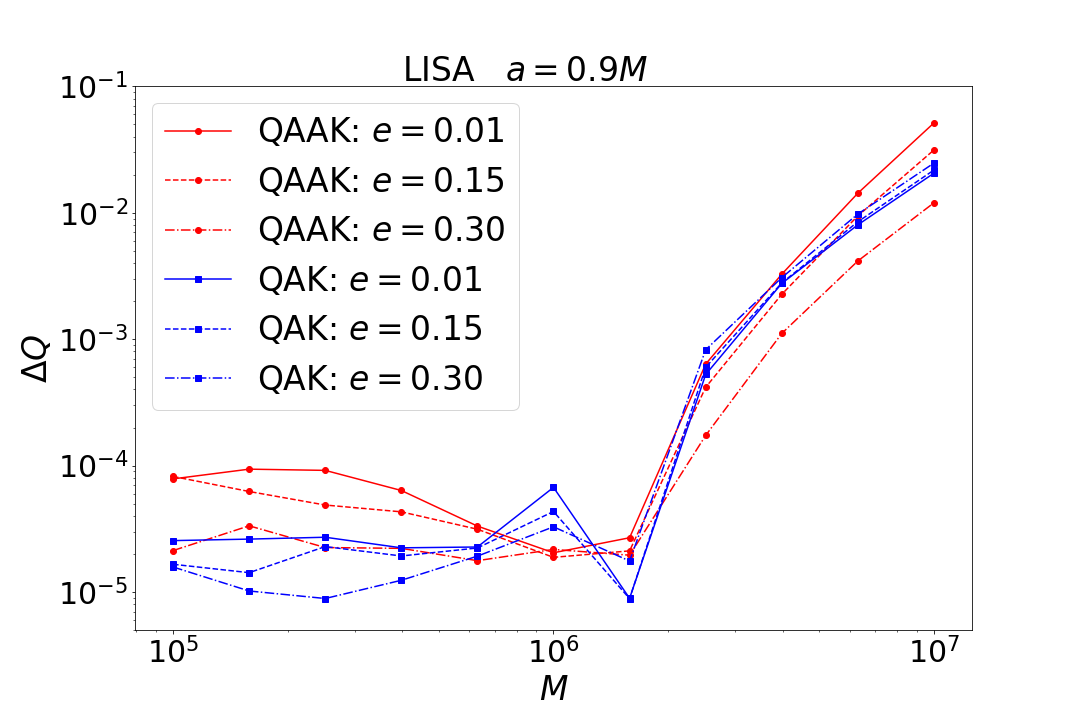}
\caption{The accuracy of $Q$ for LISA using \ac{QAK} and \ac{QAAK}.
The red curves are the result of \ac{QAK}, and the blue curves are the result of \ac{QAAK}.
Each figure for a different spin $a/M=0.3,0.6,0.9$.
The result is calculated for the \ac{MBH} with mass in the range of $10^5M_\odot\sim10^7M_\odot$.
On each figure, we choose three initial eccentricities $e=0.01,0.15,0.3$.}\label{fig:PEL}
\end{figure*}

\begin{figure*}
\centering
\includegraphics[width=0.32\textwidth]{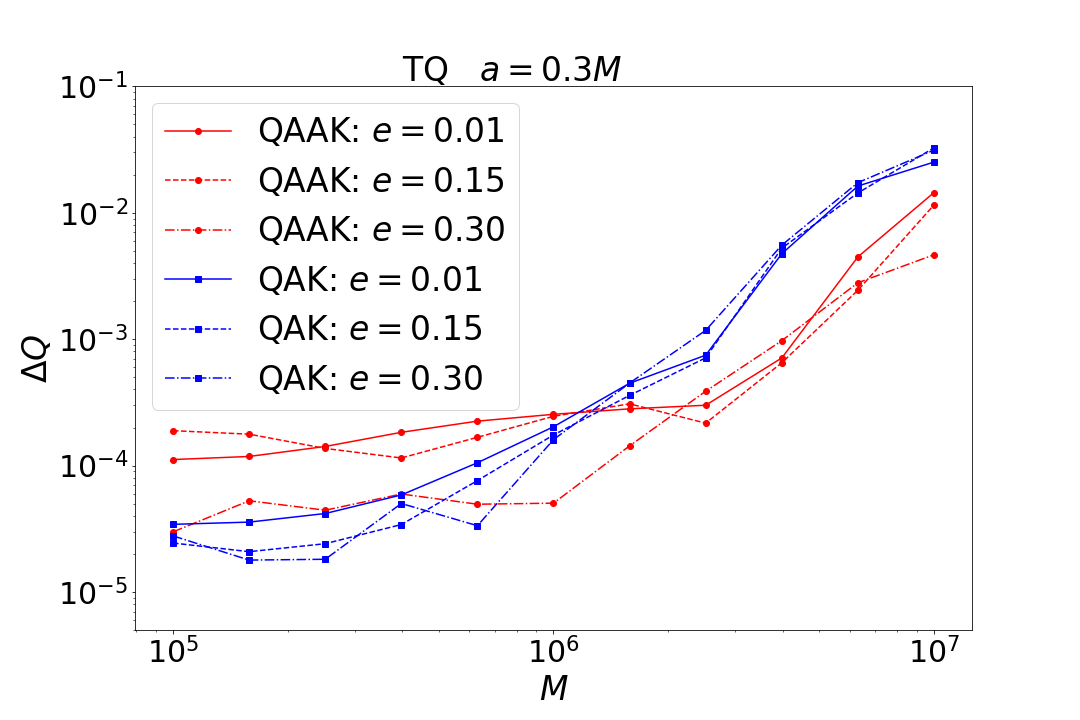}
\includegraphics[width=0.32\textwidth]{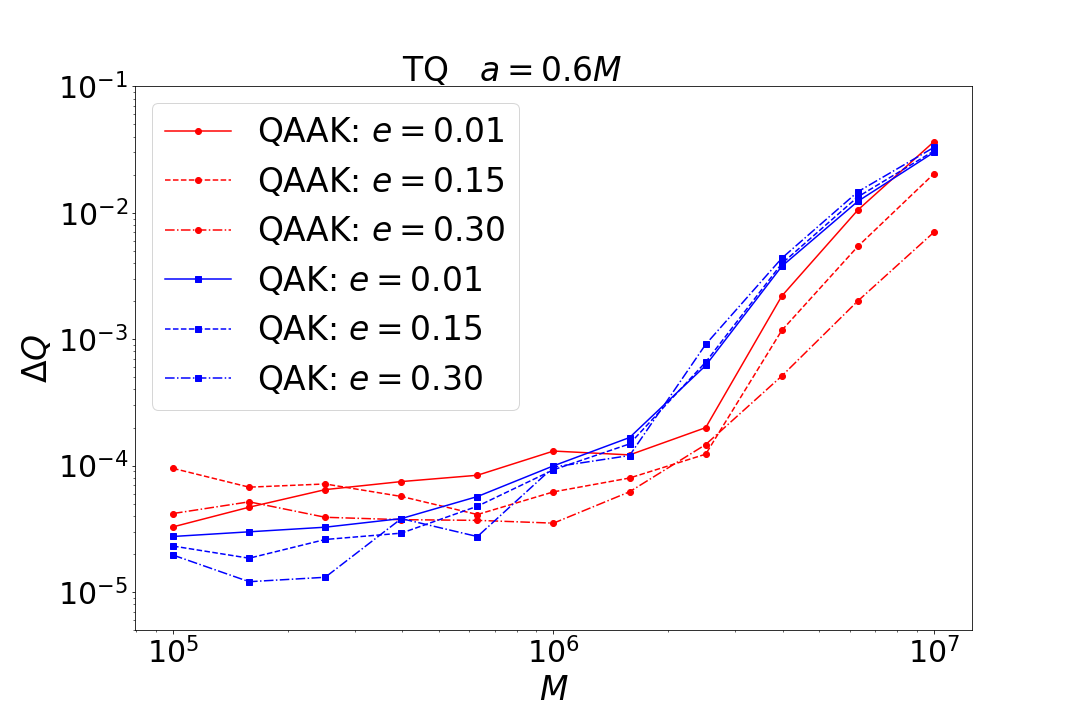}
\includegraphics[width=0.32\textwidth]{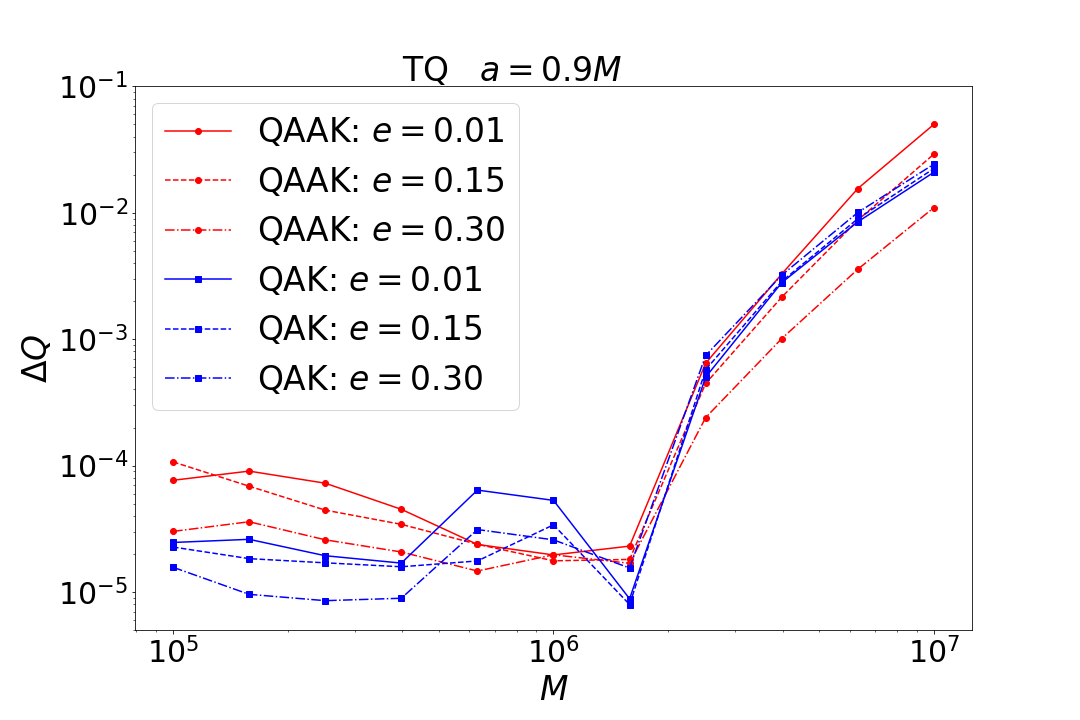}
\caption{The accuracy of $Q$ for TianQin using \ac{QAK} and \ac{QAAK}.
The red curves are the result of \ac{QAK}, and the blue curves are the result of \ac{QAAK}.
Each figure for a different spin $a/M=0.3,0.6,0.9$.
The result is calculated for the \ac{MBH} with mass in the range of $10^5M_\odot\sim10^7M_\odot$.
On each figure, we choose three initial eccentricities $e=0.01,0.15,0.3$.}\label{fig:PET}
\end{figure*}

In this paper, we considered both LISA and TianQin since their sensitivity band is a little bit different.
We compared the results of \ac{QAK} and \ac{QAAK} in FIG. \ref{fig:PEL} for LISA,
and in FIG. \ref{fig:PET} for TianQin.

The power spectral density $S_n(f)$ is chosen to be the sky averaged one for LISA \cite{Cornish:2018dyw} and TianQin \cite{Shi:2019hqa}. The length of the signal is chosen to be 1 yr.
To have a fair comparison, we normalize the \ac{SNR} to $\rho=100$.
The red shifted mass of the \ac{CO} is fixed to be $\mu=10M_{\odot}$,
and the inclination angle $\iota$ is chosen to be $\pi/3$.
The parameters $(M,a,e)$ are chosen to be different values for analyzation.
The initial value of $p$ is chosen to be $6.5M$:
for the events which plunge less than 1 yr, we will calculate backwards until the length reaches 1 yr,
for the events which will not plunge after evolve for 1 yr, we will not evolve it after that.

We can find that the \ac{QAK} model is not sensitive to the eccentricity, while in the \ac{QAAK} model,
the \ac{PE} accuracy for sources with different eccentricity will vary by several times.
Another interesting feature is that for lower spin,
the accuracy of \ac{QAAK} is higher for more massive source.
But for higher spin, this difference is not very obvious.
The reason of these feature should be the frequency evolution is different in these two waveforms.

Since we have normalized the \ac{SNR} of all the signals to be the same value,
the \ac{PE} accuracy for both detectors seems to be the same.
But if we fix the distance of each source, TianQin will have higher \ac{SNR} for sources with lower mass, and vice versa.
So for the same source, the accuracy is better for TianQin in the lower mass part,
and it's better for LISA in the higher mass part.
This meets our expectations since the band of TianQin is higher than LISA.
The \ac{MBH}s with mass in the range of $10^5M_\odot\sim10^6M_\odot$ can be measured with very high accuracy.
For the systems with mass larger than $10^6M_\odot$, the accuracy will be worse and worse.

We also listed part of the results in TABLE. \ref{LISAtable} for LISA and TABLE. \ref{TQtable} for TQ,
The first line in boldface of each mass corresponding to the result given by \ac{QAAK} waveform,
while the second line in plain face corresponding to the result given by \ac{QAK} waveform.
We also listed the result without including the estimation of $Q$ in the bracket.

\begin{table*}
\renewcommand\arraystretch{1.5}
\begin{tabular}{|c|c|c|c|c|} \hline
$M(M_{\odot})$& $\Delta(\ln\mu)$ & $\Delta(\ln M)$ & $\Delta(a/M)$ & $\Delta Q$\\
\hline

\multirow{2}{2cm}{\centering $10^5$} & $\mathbf{7.2\times 10^{-7}(6.5\times 10^{-7})}$ &
$\mathbf{1.2\times 10^{-6}(1.0\times 10^{-6})}$ & $\mathbf{3.5\times 10^{-6}(1.4\times 10^{-6})}$ & $\mathbf{2.1\times 10^{-5}(-)}$ \\
& $8.7\times 10^{-7}(6.6\times 10^{-7})$ & $1.1\times 10^{-6}(1.0\times 10^{-6})$ &
$1.3\times 10^{-6}(1.3\times 10^{-6})$ & $1.5\times 10^{-5}(-)$ \\
\hline

\multirow{2}{2cm}{\centering $10^6$} & $\mathbf{2.3\times 10^{-6}(1.6\times 10^{-6})}$ &
$\mathbf{6.9\times 10^{-7}(6.5\times 10^{-7})}$ & $\mathbf{4.3\times 10^{-6}(8.2\times 10^{-7})}$ & $\mathbf{2.2\times 10^{-5}(-)}$ \\
&$3.2\times 10^{-6}(4.8\times 10^{-7})$ & $1.0\times 10^{-6}(3.1\times 10^{-7})$ &
$5.4\times 10^{-6}(9.0\times 10^{-7})$ & $3.3\times 10^{-5}(-)$ \\
\hline

\multirow{2}{2cm}{\centering $10^7$} & $\mathbf{6.9\times 10^{-4}(4.5\times 10^{-4})}$ &
$\mathbf{2.4\times 10^{-4}(2.3\times 10^{-4})}$ & $\mathbf{2.4\times 10^{-3}(1.1\times 10^{-4})}$ & $\mathbf{1.2\times 10^{-2}(-)}$ \\
&$2.2\times 10^{-3}(2.5\times 10^{-4})$ & $1.7\times 10^{-4}(1.7\times 10^{-4})$ &
$4.3\times 10^{-3}(3.8\times 10^{-5})$ & $2.5\times 10^{-2}(-)$ \\
\hline

\end{tabular}
\caption{The result of QAAK and QAK for LISA for $e=0.3$, $a=0.9M$ and the mass of \ac{MBH} is chosen to be different values.
The boldfaced data in the first line for each mass corresponding to the \ac{PE} accuracy given by the \ac{QAAK},
while the plain-faced  in the second line for each mass is given by \ac{QAK}.
The data in the bracket is the \ac{PE} result without including the estimation of $Q$.}
\label{LISAtable}
\end{table*}

\begin{table*}
\renewcommand\arraystretch{1.5}
\begin{tabular}{|c|c|c|c|c|} \hline
$M(M_{\odot})$ & $\Delta(\ln\mu)$ & $\Delta(\ln M)$ & $\Delta(a/M)$ & $\Delta Q$\\
\hline

\multirow{2}{2cm}{\centering $10^5$} & $\mathbf{1.0\times 10^{-6}(8.5\times 10^{-7})}$ &
$\mathbf{1.5\times 10^{-6}(1.1\times 10^{-6})}$ & $\mathbf{5.4\times 10^{-6}(1.9\times 10^{-6})}$ & $\mathbf{3.0\times 10^{-5}(-)}$ \\
& $1.0\times 10^{-6}(6.6\times 10^{-7})$ & $1.2\times 10^{-6}(1.1\times 10^{-6})$ &
$1.7\times 10^{-6}(1.6\times 10^{-6})$ & $1.6\times 10^{-5}(-)$ \\
\hline

\multirow{2}{2cm}{\centering $10^6$} & $\mathbf{2.3\times 10^{-6}(1.5\times 10^{-6})}$ &
$\mathbf{6.0\times 10^{-7}(5.6\times 10^{-7})}$ & $\mathbf{4.0\times 10^{-6}(8.0\times 10^{-7})}$ & $\mathbf{2.0\times 10^{-5}(-)}$ \\
&$2.6\times 10^{-6}(5.0\times 10^{-7})$ & $8.1\times 10^{-7}(2.6\times 10^{-7})$ &
$4.3\times 10^{-6}(9.3\times 10^{-7})$ & $2.6\times 10^{-5}(-)$ \\
\hline

\multirow{2}{2cm}{\centering $10^7$} & $\mathbf{6.7\times 10^{-4}(4.2\times 10^{-4})}$ &
$\mathbf{2.4\times 10^{-4}(2.1\times 10^{-4})}$ & $\mathbf{2.1\times 10^{-3}(1.1\times 10^{-4})}$ & $\mathbf{1.1\times 10^{-2}(-)}$ \\
&$2.4\times 10^{-3}(2.2\times 10^{-4})$ & $1.4\times 10^{-4}(1.4\times 10^{-4})$ &
$4.4\times 10^{-3}(3.2\times 10^{-5})$ & $2.4\times 10^{-2}(-)$ \\
\hline

\end{tabular}
\caption{The result of QAAK and QAK for TianQin for $e=0.3$, $a=0.9M$ and the mass of \ac{MBH} is chosen to be different values.
The boldfaced data in the first line for each mass corresponding to the \ac{PE} accuracy given by the \ac{QAAK},
while the plain-faced  in the second line for each mass is given by \ac{QAK}.
The data in the bracket is the \ac{PE} result without including the estimation of $Q$.}
\label{TQtable}
\end{table*}

We can find that the result of \ac{QAK} and \ac{QAAK} almost agree with each other at the same order of magnitude.
But since \ac{QAAK} is developed based on the \ac{AAK} waveform with higher accuracy,
the waveform should be more reliable in the realistic matched filtering.

\section{conclusion}\label{sec:conclusion}

Based on the \ac{AAK} model, we considered the quadrupole moment corrections due to the off-Kerr deviations in the geometry of massive black holes.
We modified the evolution equations of the orbital frequencies in the \ac{AK} side,
as well as the constants' evolution equations in the \ac{NK} side,
by a simple substitution of $a^2/M^2$ with $-\mathcal{Q}/M^3$.
The fundamental frequencies are obtained in the bumpy Kerr \ac{BH} spacetime with arbitrary quadrupole moment.
The definition of the constants and the geodesic equations are still the one for Kerr \ac{BH}.

We also compare \ac{QAAK} and \ac{QAK}'s ability of \ac{PE} for various sources with LISA and TianQin.
We find that although these waveforms are expected to have a different accuracy,
the \ac{PE} accuracy will be influenced only for quite a few times differences.
Hopefully, with a more accurate waveform, \ac{QAAK} may be used in data analysis for testing the no-hair theorem with \ac{EMRI}.

\acknowledgments

We thank Alvin J. K. Chua and Yiming Hu for helpful discussion.
This work makes use of the Black Hole Perturbation Toolkit.
The code of this work is developed based on the EMRI Kludge Suite from https://github.com/alvincjk/EMRI\_Kludge\_Suite.
This work has been supported by the Natural Science Foundation of China (Grant Nos.11805286).

\bibliographystyle{unsrt}
\bibliography{QAAK}

\end{document}